\documentclass[reprint,superscriptaddress,nofootinbib,amsmath,amssymb,aps]{revtex4-2}
\usepackage{graphicx}
\usepackage{dcolumn}
\usepackage{bm}
\usepackage{hyperref}
\hypersetup{colorlinks,citecolor=blue,linkcolor=blue,urlcolor=blue,}

\begin{document}

\title{Constraints on Self-Interacting Fuzzy Dark Matter from the Stellar Kinematics of the Dwarf Galaxy Leo II}
\author{Yi Zhao}
\email{zhaoyi@tjnu.edu.cn}
 \affiliation{College of Physics and Materials Science, Tianjin Normal University, Tianjin 300387, China}
\author{Yu-Ming Yang}
\email{yangyuming@ihep.ac.cn}
 \affiliation{%
 State Key Laboratory of Particle Astrophysics, Institute of High Energy Physics, Chinese Academy of Sciences, Beijing 100049, China}
\affiliation{
 School of Physical Sciences, University of Chinese Academy of Sciences, Beijing 100049, China 
}%
\author{Xiao-Jun Bi}
\email{bixj@ihep.ac.cn}
\affiliation{%
 State Key Laboratory of Particle Astrophysics, Institute of High Energy Physics, Chinese Academy of Sciences, Beijing 100049, China}
\affiliation{
 School of Physical Sciences, University of Chinese Academy of Sciences, Beijing 100049, China 
}%
\author{Peng-Fei Yin}
\email{yinpf@ihep.ac.cn}
\affiliation{%
 State Key Laboratory of Particle Astrophysics, Institute of High Energy Physics, Chinese Academy of Sciences, Beijing 100049, China}

\begin{abstract}
The one-parameter fuzzy dark matter (FDM) model has faced increasingly stringent constraints from both Lyman-$\alpha$ forest observations and local measurements of dwarf galaxies. A natural extension to mitigate these limits is the inclusion of FDM self-interactions. In this study, we derive constraints in the two-dimensional parameter space $(m_a, f_a)$ using the dark matter density profile inferred from a Jeans analysis of the stellar kinematics in the dwarf galaxy Leo II, which has previously been employed to constrain non-interacting FDM. We find that, for a fixed particle mass $m_a$, attractive (repulsive) self-interaction leads to a more concentrated (more diffuse) FDM density profile relative to the non-interacting case, thereby improving (worsening) agreement with the Jeans analysis results. Our results indicate that, for either attractive or repulsive SI with strength $f_a^{-1}\lesssim 10^{-14}\,\mathrm{GeV}^{-1}$, the $95\%$ confidence-level lower limits on $m_a$ lies within the range $(1-10)\times10^{-22}\,\mathrm{eV}$, although the precise bounds depend to some extent on the statistical method employed. This analysis simultaneously constrains the two parameters $(m_a, f_a)$ without relying on assumptions about cosmological or galaxy evolution histories, and thus offers a complementary probe to existing constraints.
\end{abstract}


\maketitle
\section{Introduction}
As a widely studied alternative to the cold dark matter (CDM) paradigm, fuzzy dark matter (FDM) \cite{hu2000fuzzy, peebles2000fluid, hui2017ultralight, hui2021wave, eberhardt2025ultralightfuzzydarkmatter, schive2025fuzzydarkmattersimulations} has attracted considerable attention in recent years and offers potential avenues for alleviating several small-scale challenges faced by CDM \cite{Bullock:2017xww}. For instance, the soliton core that forms at the center of FDM halos, resulting from the balance between quantum pressure and gravity, may provide a possible explanation for the core-cusp problem \cite{de2010core}. Moreover, the suppression of dynamical friction in FDM has been proposed as a solution to the rapid orbital decay problem of globular clusters in Fornax \cite{Lora:2011yc, Lancaster:2019mde}, while dynamical heating induced by wave interference has been suggested as a mechanism to explain the extremely diffuse stellar distributions observed in some dwarf galaxies \cite{Montes:2023ahn, Yang:2024vgw, Yang:2024ixt}.

An appealing feature of the simplest FDM model is that its dependence on a single fundamental parameter, the particle mass $m_a$. However, precisely because the model has only one free parameter, it has become subject to increasingly stringent constraints. These include bounds from large-scale cosmological observations, such as the Lyman-$\alpha$ forest \cite{Rogers:2020ltq}, as well as constraints from local dwarf galaxy observations. For example, several studies have used stellar distributions and velocity dispersions in dwarf galaxies to place strong limits on $m_a$ through the dynamical heating effect \cite{Marsh:2018zyw, Chiang:2021uvt, Dalal:2022rmp, Teodori:2025rul, May:2025ppj}. Furthermore, many works \cite{Hayashi:2021xxu, DeMartino:2023cgg, Benito:2025xuh} have constrained $m_a$ via Jeans analyses based on the parametric soliton density profile given by cosmological simulations \cite{Schive:2014dra, Schive:2014hza, Chan:2025hhg, Schwabe:2021jne}. More recently, an alternative approach to traditional Jeans analyses has been proposed and applied to stellar kinematic data of the dwarf galaxy Leo II \cite{Zimmermann:2024xvd}. This method yields a very stringent constraint on $m_a$ and does not rely on assumptions regarding cosmological or galaxy evolution histories.

A natural extension of the FDM framework involves introducing a self-interaction (SI) term of the FDM fields with a coupling $f_a$. This extension increases the model's parametric freedom, gives rise to richer phenomenology \cite{Chavanis_2011, Chavanis_2011_2, Levkov_2017, Helfer_2017, Ferreira_2021, Glennon_2022, Mocz_2023, painter2024attractivemodelsimulatingfuzzy, Glennon_2024, Berezhiani:2025maf}, and may thereby open a broader region of parameter space that can evade the aforementioned constraints. From the perspective of particle physics, SIs of (pseudo-)scalar particles, such as the QCD axion \cite{Di_Luzio_2020} and axion-like particles (ALPs) \cite{Marsh_2016, OHare:2024nmr}, can arise naturally from the periodic potential. Consequently, self-interacting fuzzy dark matter (SIFDM) has attracted growing attention in recent years. However, systematic and robust constraints on the SIFDM parameter space $(m_a, f_a)$ remain relatively limited.

In this work, we apply the methodology introduced in Ref. \cite{Zimmermann:2024xvd} to constrain the parameter space of SIFDM using stellar kinematic observations of the dwarf galaxy Leo II. The procedure consists of three main steps. First, we perform a Jeans analysis of Leo II using the general coreNFWtides \cite{Read:2018pft} parameterization for the DM density profile, thereby obtaining the allowed range of such profiles. We then take the median profile of their distribution as the fiducial input and reconstruct the corresponding SIFDM wavefunction via the eigenstate decomposition method \cite{Lin_2018, Yavetz_2022}. This reconstruction is achieved by fitting the input profile through appropriate weighting of the quantum eigenstates.  Finally, we introduce three statistical methods to quantify the consistency between the density profile derived from the reconstructed SIFDM profile with a given parameter set $(m_a,f_a)$ and the density profile directly inferred from the Jeans analysis, thereby deriving $95\%$ confidence-level constraints on $(m_a,f_a)$.
An attractive (repulsive) SI makes the SIFDM density profile more concentrated (more diffuse). Consequently, for a fixed particle mass $m_a$, a stronger attractive (repulsive) SI strength $f_a^{-1}$ leads to better (worse) agreement with the results of the Jeans analysis. Similarly, for a fixed SI strength $f_a^{-1}$, increasing (decreasing) the particle mass $m_a$ improves (worsens) the agreement. 


The remainder of this paper is organized as follows. Sec. \ref{Sec2} introduces the observational data of Leo II and our Jeans analysis. Sec. \ref{Sec3} presents the construction of the SIFDM halo wavefunction using the eigenstate decomposition method. Sec. \ref{Sec4} describes the statistical methods and the resulting constraints on the parameter space. Finally, Sec. \ref{Sec5} summarizes our conclusions.

\section{Jeans analysis\label{Sec2}}
\subsection{Method}
The dwarf spheroidal galaxy Leo II, a  classical satellite of the Milky Way, provides a valuable astrophysical laboratory for investigating DM through stellar kinematics. It lies at a heliocentric distance of approximately 233 kpc. A large spectroscopic survey of red giant stars in the direction of Leo II has yielded a well-defined kinematic data set~\cite{2017ApJ...836..202S,2017AJ....153..254S}. From a total of 336 observed stars, 175 are identified as members based on their radial velocities and surface gravities. This membership selection typically relies on a velocity window $\sim 57 \lesssim v \lesssim 100~\mathrm{km\,s^{-1}}$ together with constraints on stellar atmospheric parameters, ensuring a clean sample with minimal contamination. The resulting member sample exhibits a systemic velocity of 
$\sim 78.5~\mathrm{km\,s^{-1}}$ and a velocity dispersion of $\sim 7.4~\mathrm{km\,s^{-1}}$. The data show no significant evidence for rotation, tidal disruption, or kinematic substructure. Importantly, its velocity dispersion profile is consistent with being flat out to large radii, suggesting that Leo II is embedded in an extended DM halo. This high-quality kinematic data set, combined with the relatively large number of confirmed member stars and repeated measurements for many of them, makes Leo II particularly suitable for dynamical modeling via the Jeans analysis.

The dynamical behavior of stars embedded in a gravitational potential
can be described by the Jeans equation, which is derived from the
collisionless Boltzmann equation through appropriate moment integrations.
Under assumptions of spherical symmetry, dynamical equilibrium, and negligible
rotational support, the second-order Jeans equation reduces to
\begin{equation}
\frac{1}{\nu(r)}
\frac{d}{dr}
\left[
\nu(r)\sigma_r^2(r)
\right]
+
\frac{2\beta}{r}
\sigma_r^2(r)
=
-
\frac{G M(<r)}{r^2},
\label{eq:jeans}
\end{equation}
where $\nu(r)$ denotes the three-dimensional stellar density profile,
$\sigma_r(r)$ is the radial velocity dispersion, $G$ is the gravitational
constant, and $M(<r)$ represents the enclosed mass within radius $r$.
In dwarf spheroidal galaxies, the stellar component contributes only a
small fraction of the total mass. consequently, it is a valid approximation to replace the total mass in Eq.~\eqref{eq:jeans} with the DM mass alone. 
The parameter $\beta$ characterizes the velocity anisotropy of the
stellar system, quantifying the relative importance of tangential and
radial motions, and is defined as $\beta=1-\sigma_\theta^2/\sigma_r^2$. In this study,  we adopt a constant velocity anisotropy $\beta$ for simplicity.

The general solution to Eq.~\eqref{eq:jeans} can be formally expressed as
\begin{equation}
\nu(r)\sigma_r^2(r)
=
\frac{1}{A(r)}
\int_r^\infty
A(s)
\nu(s)
\frac{G M(<s)}{s^2}
\, ds,
\label{eq:nu_sigma_r}
\end{equation}
where $A(r)=A_{r_1}\exp[\int_{r_1}^r\frac{2}{t}\beta_{ani}(t)dt]$
is an integrating factor that depends on the anisotropy profile.
The reference radius $r_1$ is arbitrary and introduces only an overall
normalization factor, which cancels out in Eq.~\eqref{eq:nu_sigma_r}
\cite{2015MNRAS.446.3002B}.
Since astronomical observations provide only projected quantities,
namely the surface number density and the line-of-sight velocity
dispersion, it is necessary to project the three-dimensional solution
onto the plane of the sky. This is achieved via the Abel transform,
leading to the projected velocity dispersion profile
\begin{equation}
\sigma_p^2(R)
=
\frac{2}{I(R)}
\int_R^\infty
\left[
1 - \beta \frac{R^2}{r^2}
\right]
\frac{\nu(r)\sigma_r^2(r)\, r}
{\sqrt{r^2 - R^2}}
\, dr,
\label{eq:sigma_p}
\end{equation}
where $\sigma_p^2(R)$ denotes the line-of-sight velocity dispersion at
projected radius $R$, and $I(R)$ is the projected surface brightness
profile.

We model the surface brightness using an exponential profile $I(R)=I_0exp(-R/r_e)$,
which provides a good description for many dwarf spheroidal galaxies.
The corresponding three-dimensional stellar density profile is obtained
through an inverse Abel transform and is given by $\nu(r)=I_0/(\pi r_e)\times K_0(r/r_e)$,
where $K_0$ is the modified Bessel function of the second kind of order
zero. The exponential scale length $r_e$ is related to the half-light
radius via $r_h=1.68r_e$ \cite{2018ApJ...860...66M}.

The theoretical prediction of $\sigma_p^2(R)$ is compared with the
observed stellar kinematic data through a Gaussian likelihood function.
In this work, we adopt an unbinned likelihood analysis, which avoids
information loss associated with binning procedures and is particularly
advantageous for systems with limited sample sizes. The likelihood function is defined as 
\begin{equation}
\mathcal{L}_\mathrm{dsph}=\prod_{i=1}^{N_\mathrm{stars}} \frac{\exp[-\frac{1}{2}
(\frac{(v_i-\bar{v})^2}{\sigma_p^2(R_i)+\Delta_{v_i}^2})]}{\sqrt{2\pi[\sigma_p^2(R_i)+\Delta_{v_i}^2]}},
\label{eq:likelihood}
\end{equation}
where $\sigma_p^2(R_i)$ is evaluated from Eq.~\eqref{eq:sigma_p},
$v_i$ is the line-of-sight velocity of the $i$-th star, and
$\Delta_{v_i}$ denotes the corresponding measurement uncertainty.
The observed velocities are assumed to follow a Gaussian distribution
with mean velocity $\bar{v}$.

We use kinematic data for the 175 member stars in Leo II~\cite{2017ApJ...836..202S}.
The posterior distributions of the model parameters are sampled using
the public \texttt{emcee} package.
The inferred mass profiles primarily reflect
the underlying DM distribution.


We first perform a traditional Jeans analysis, adopting an FDM density profile motivated by cosmological simulations, to constrain the FDM particle mass $m_a$ within the single-parameter FDM framework. Using the method in \cite{Hayashi:2021xxu} and an MCMC sampling with $1.035\times 10^6$ samples, we obtain a posterior distribution characterized by a median value $m_a=10^{-19.8}$ eV and a $68\%$ credible interval spanning $10^{-20.7}$ to $10^{-17.2}$ eV. See App. \ref{App_A} for details. The relatively broad interval reflects the combined influence of the limited kinematic data and intrinsic degeneracies in dynamical modeling. The inferred constraint tends to favor relatively large values of $m_a$, implying a comparatively small soliton core in Leo II. Nevertheless, within the allowed parameter space, lower values of $m_a$ remain viable, in which case a more extended soliton core can still arise and contribute non-negligibly to the inner density structure.



\subsection{coreNFWt profile}

In the CDM scenario, the DM density distribution in dwarf galaxies can be modeled by the so-called coreNFW
profile~\cite{2016MNRAS.459.2573R}. This phenomenological description captures DM halos with central cores, as expected from stellar feedback effects. Such profiles are
also used to characterize halos in alternative DM scenarios,
such as self-interacting DM~\cite{2017MNRAS.470.1542S}.

The coreNFW density profile is given by
\begin{equation}
\rho_{\rm cNFW}(r)
=
f^n\rho_{\rm NFW}(r)
+\frac{nf^{n-1}(1-f^2)}{4\pi r^2 r_c}M_{\rm NFW}(r),
\label{eq:rho_cNFW}
\end{equation}
where $\rho_{\rm NFW}(r)=\rho_s/[(r/r_s)(1+r/r_s)^2]$ is the standard NFW profile, 
and the transition function is defined as $f^n = \left[\tanh(r/r_c)\right]^n$. 
Here $r_c$ is the core radius and $n$ controls the sharpness of the transition.

The NFW scale density $\rho_s$ and scale radius $r_s$ are related to
the virial properties of the halo via
\begin{equation}
\rho_s=\frac{\rho_{\rm crit}\Delta c_{200}^3 g_c}{3}, \;\;\; r_s = \frac{r_{200}}{c_{200}},
\label{eq:rho_s}
\end{equation}
where the concentration-dependent factor is
\begin{equation}
g_c = \frac{1}{{\rm ln}(1+c_{200})-\frac{c_{200}}{1+c_{200}}},
\label{eq:g_c}
\end{equation}
and the virial radius is defined as
\begin{equation}
r_{200}=\left(\frac{3 M_{200}}{4\pi \Delta \rho_{\rm crit}}\right)^{1/3}.
\label{eq:r_200}
\end{equation}
Here $M_{200}$ is the mass enclosed within $r_{200}$, inside which the
mean density is $\Delta=200$ times the critical density $\rho_{\rm crit}$,
and $c_{200}$ is the halo concentration parameter.

The cumulative mass profile of the coreNFW model is given by
\begin{equation}
M_{\rm cNFW}(r) = M_{\rm NFW}(r)f^n,
\label{eq:mass_cNFW}
\end{equation}
where the NFW enclosed mass is
\begin{equation}
M_{\rm NFW}(r) = M_{\rm 200}g_c\left[{\rm ln}\left(1+\frac{r}{r_s}\right)-\frac{r/r_s}{1+r/r_s}\right].
\label{eq:msss_NFW}
\end{equation}

Furthermore, satellite galaxies may be subject to tidal stripping by their
host halo, which can steepen the outer density profile. To account for
this effect, we adopt a modified coreNFW model,
defined as
\begin{equation}
\rho_{\text{cNFWt}}(r) =
\begin{cases}
\rho_{\text{cNFW}}(r), & r < r_t, \\
\rho_{\text{cNFW}}(r_t) \left( \frac{r}{r_t} \right)^{-\delta}, & r > r_t,
\end{cases}
\label{eq:rho_cNFWt}
\end{equation}
where $r_t$ denotes the tidal truncation radius, beyond which the density
profile follows a power-law with logarithmic slope $\delta$. The enclosed mass profile is modified to
\begin{equation}
\begin{split}
M_{\rm cNFWt}(<r)
=
\begin{cases}
M_{\rm cNFW}(<r), & r \le r_t, \\[6pt]
M_{\rm cNFW}(r_t) +
\displaystyle
\frac{4\pi \rho_{\rm cNFW}(r_t) r_t^3}{3-\delta} \\
\times \left[
\left(\frac{r}{r_t}\right)^{3-\delta}
- 1
\right],
& r > r_t,
\end{cases}
\end{split}
\label{eq:mass_truncated}
\end{equation}
which ensures continuity of the mass profile at $r=r_t$.

In our analysis, the density profile parameters $M_{200}$, $c_{200}$, $r_c$, $n$,
$\delta$, $r_t$, and the velocity anisotropy $\beta$ are treated as
free parameters in the MCMC sampling.
We adopt the following prior ranges:
$7<{\rm log}_{10}(M_{200}/{\rm M_{\odot}})<9$,
$1.0<{\rm log_{10}}(c_{200})<1.7$,
$-1.7<{\rm log_{10}}(r_c/{\rm kpc})<0.3$,
$0<n<1$, $3.5<\delta<5$,
$0.2<{\rm log_{10}}(r_t/r_h)<1.2$, and $-5<\beta<1$.
We impose the physical condition $r_c<r_t$ to ensure that the core
region lies within the tidal radius.
These priors are chosen to encompass a  physically plausible parameter
space while remaining sufficiently broad to avoid biasing the results.

\begin{figure}[htbp]
    \centering    
    \includegraphics[width=\linewidth]{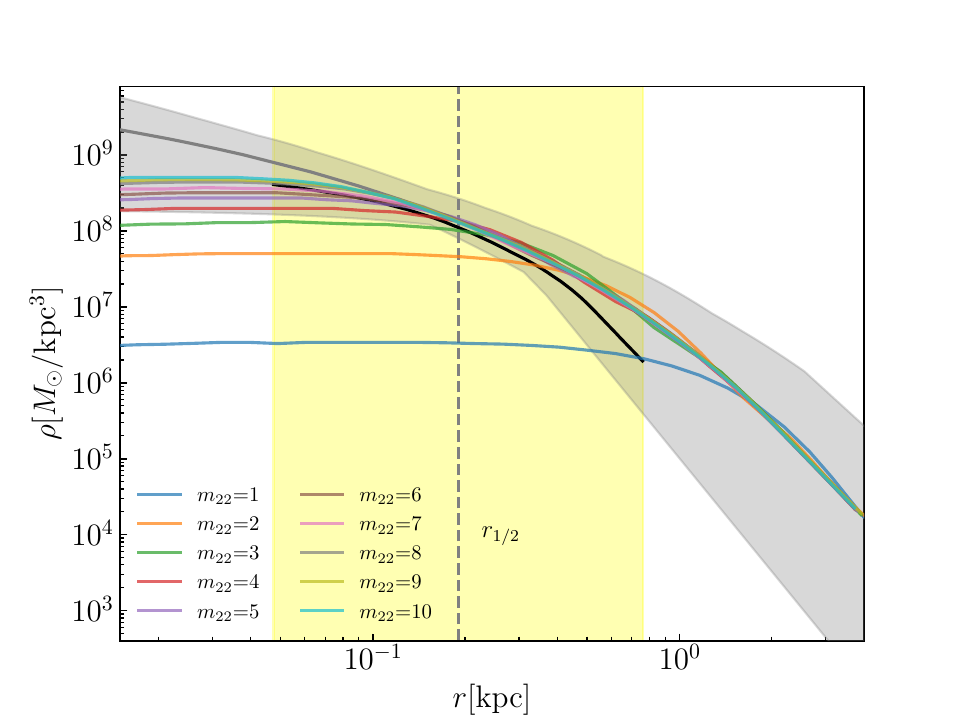}
    \includegraphics[width=\linewidth]{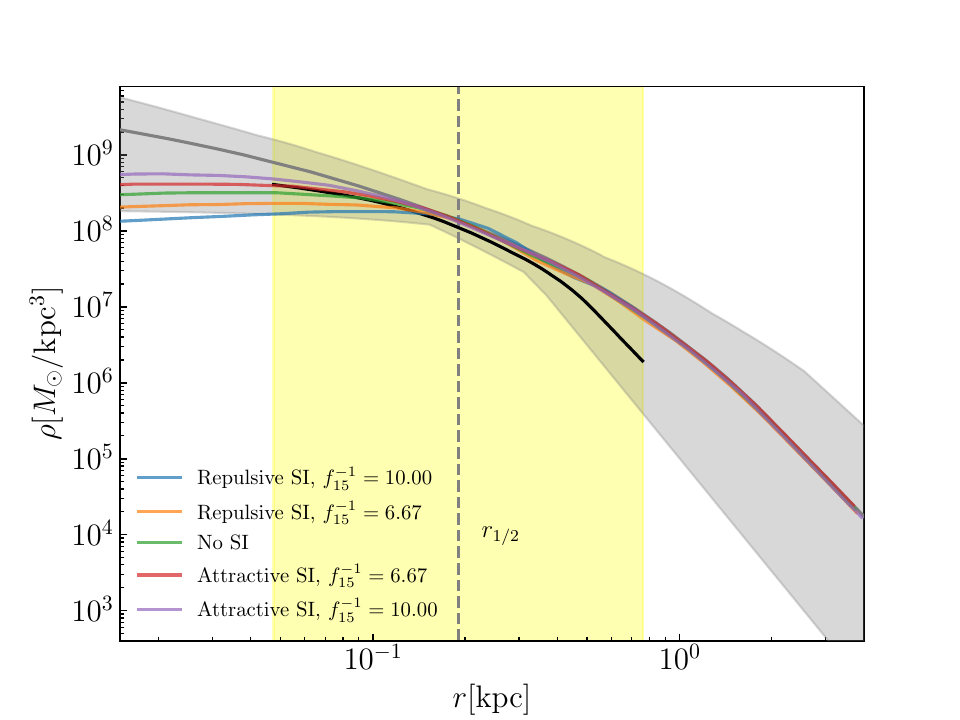}
    \caption{Comparison between the constructed FDM density profiles $\rho_\mathrm{out}$ and the density profiles inferred from the Jeans analysis. The upper panel shows the non-interacting case for different values of $m_{22}$. The lower panel corresponds to a fixed $m_{22}=6$ with varying signs and strengths of the SI. In both panels, the gray shaded region represents the spread of the $1.596 \times 10^5$ density profiles obtained from the Jeans analysis. The gray solid line denotes their median. The gray dashed line marks the half-light radius of Leo II $r_{1/2} \sim 190$ pc. The yellow shaded region indicates the radial interval $[r_{1/2},4r_{1/2}]$ used to set constraints in one method. Within this interval, the solid black curve connects the points corresponding to the lowest $5\%$ of the density distributions from the Jeans analysis.}
    \label{fit}
\end{figure}

The MCMC configuration employs 28 walkers, 
each evolved for a total of 6000 steps. To ensure chain convergence and mitigate the influence of initial conditions, the first 300 steps of each walker are discarded as burn-in, yielding a final ensemble of 159,600 posterior samples.
The inferred posterior distributions of the model parameters are
${\rm log}_{10}(M_{200}/{\rm M_{\odot}})=8.5_{-0.3}^{+0.3}$,
${\rm log_{10}}(c_{200})=1.5_{-0.1}^{+0.1}$,
${\rm log_{10}}(r_c/{\rm kpc})=-1.0_{-0.5}^{+0.6}$,
$n=0.4_{-0.3}^{+0.4}$,
$\delta=4.2_{-0.5}^{+0.5}$,
${\rm log_{10}}(r_t/{\rm kpc})=0.04_{-0.3}^{+0.3}$,
and $\beta=-0.05_{-0.5}^{+0.3}$. 
These posterior samples are mapped into 
a corresponding set of density profiles, whose statistical distribution 
is visualized as the gray shaded region in Fig. \ref{fit}.

\section{Construction of the halo wavefunction\label{Sec3}}

The SIFDM density is determined by the wave function $\psi$ via the relation $\rho = m_a|\psi|^2$. Given a density profile, the corresponding SIFDM wavefunction that satisfies the equation of motion can be reconstructed. In this section, we describe the methodology for this construction.

\subsection{Theoretical background}
SIFDM consists of a large number of ultralight bosons. Since the occupation number within a de Broglie volume is much larger than unity, quantum fluctuations are negligible, and the system can be well described by a classical real scalar field $\phi$. In natural units with $\hbar=c=1$, the corresponding Lagrangian is given by \cite{Ferreira_2021}
\begin{equation}
    \mathcal{L}=-\frac{1}{2}g^{\mu\nu}\partial_\mu\phi\partial_\nu\phi-\frac{1}{2}m_a^2\phi^2-\frac{\lambda}{4!}\phi^4,
\end{equation}
where we include a $\phi^4$ SI term with a  coupling constant $\lambda=\pm m_a^2/f_a^2$. For a fixed particle mass $m_a$, the interaction strength is therefore controlled by $f_a^{-1}$, with the positive and negative signs corresponding to repulsive and attractive SI, respectively. In axion or ALP theories \cite{Marsh_2016, OHare:2024nmr}, such a SI term can naturally arise from the expansion of the periodic axion potential $V(\phi)=m_a^2f_a^2\left[1-\cos(\phi/f_a)\right]\simeq m_a^2\phi^2/2-(1/4!)(m_a^2/f_a^2)\phi^4+\cdots$, which corresponds to an attractive SI. 

In the nonrelativistic limit, SIFDM is typically described by a complex wavefunction $\psi$, which is related to the real scalar field $\phi$ via $\phi=(1/\sqrt{2m_a})(\psi e^{-im_at}+\psi^\star e^{im_at})$. The equation of motion for $\psi$ then takes the form of the Gross-Pitaevskii-Poisson equation \cite{Ferreira_2021, painter2024attractivemodelsimulatingfuzzy}
\begin{equation}
    \begin{aligned}
        &i\hbar\frac{\partial\psi}{\partial t}=-\frac{\hbar^2}{2m_a}\nabla^2\psi+m_a\Phi\psi\pm\frac{\hbar^3c^3}{8f_a^2}|\psi|^2\psi,\\
        &\nabla^2\Phi=4\pi G\rho.
    \end{aligned}
\end{equation}
Here we explicitly restore $\hbar$ and $c$ for numerical convenience, and neglect the effect of cosmic expansion, which is a good approximation on galactic scales. The gravitational potential $\Phi$ satisfies the Poisson equation, where the SIFDM density is given by $\rho=m_a|\psi|^2$. Since the system considered here is DM dominated, the gravitational potential contributed by stars can be neglected, and $\Phi$ is therefore determined solely by the SIFDM density.

\subsection{Eigenstate decomposition}
To derive the FDM wave function from a given profile, we employ the eigenstate decomposition method \cite{Lin_2018, Yavetz_2022}, in which the wavefunction is expressed as a linear combination of eigenstates,
\begin{equation}
    \psi(t_0,\boldsymbol{x})=\sum_{nlm}a_{nlm}\Psi_{nlm}(\boldsymbol{x}).
\end{equation}
Here, $n$, $l$, and $m$ denote the principal, angular, and magnetic quantum numbers labeling the eigenstates $\Psi_{nlm}$, while the expansion coefficients $a_{nlm}$ are complex numbers. 

The eigenstates $\Psi_{nlm}$ are obtained as solutions of the time-independent Gross-Pitaevskii equation,
\begin{equation}
\begin{aligned}
      -\frac{\hbar^2}{2m_a}\nabla^2\Psi_{nlm}(\boldsymbol{x})&+m_a\Phi_\mathrm{in}\Psi_{nlm}(\boldsymbol{x})\\
      &\pm\frac{\hbar^3c^3\rho_\mathrm{in}}{8m_af_a^2}\Psi_{nlm}(\boldsymbol{x})=E_{nl}\Psi_{nlm}(\boldsymbol{x}).  
\end{aligned}
\end{equation}
Here, we adopt an effective approximation to linearize the equation by fixing the interaction potential and the gravitational potential to quantities that are independent of the actual solution $\psi$. Since our goal is to reconstruct the result from the Jeans analysis, we take $\rho_{\mathrm{in}}$ to be the median density profile obtained from the Jeans analysis represented by the gray solid lines in Fig. \ref{fit}. The corresponding potential $\Phi_{\mathrm{in}}$ is then determined from $\rho_{\mathrm{in}}$ through the Poisson equation. As a result, both $\rho_{\mathrm{in}}$ and $\Phi_{\mathrm{in}}$ are spherically symmetric, which allows the eigenstates to be separated into a product of a radial wavefunction and spherical harmonics,
\begin{equation}
    \Psi_{nlm}(\boldsymbol{x})=R_{nl}(r)Y_l^m(\theta,\phi).
\end{equation}
The radial wavefunction can be obtained by solving for $u_{nl}(r)\equiv rR_{nl}(r)$, which satisfies the following differential equation
\begin{equation}
\begin{aligned}
    -\frac{\hbar^2}{2m_a}\frac{d^2u_{nl}}{dr^2}+&\left[\frac{\hbar^2}{2m_a}\frac{l(l+1)}{r^2}+m_a\Phi_\mathrm{in}(r)\right.\\
    &\qquad\qquad\left.\pm\frac{\hbar^3c^3\rho_\mathrm{in}(r)}{8m_af_a^2}\right]u_{nl}=E_{nl}u_{nl}.
\end{aligned}
\end{equation}
We solve this eigenvalue problem using the shooting method, obtaining a series of $u_{nl}$ and the corresponding eigenenergies $E_{nl}$. The numerical procedure is similar to that described in Ref.~\cite{Yang:2024ixt}, with the addition of the self-interaction term, which was not included previously.

The determination of the expansion coefficients $a_{nlm}=|a_{nlm}|e^{i\phi_{nlm}}$ proceeds in two steps. First, for each set of quantum numbers $(n,l,m)$, we randomly assign a phase $\phi_{nlm}$ within the range $[0,2\pi]$. The magnitudes of the coefficients, $|a_{nlm}|$, which determine the relative contributions of different eigenstates, are obtained by fitting the input profile $\rho_{\mathrm{in}}(r)$ to the random-phase-averaged (equivalently, time-averaged) density profile $\rho_{\mathrm{out}}(r)=(m_a/4\pi)\sum_{nl}(2l+1)|a_{nlm}|^2|R_{nl}(r)|^2$. To simplify the calculation, we neglect the dependence of $|a_{nlm}|$ on $m$. Under this approximation, the fitting problem reduces to minimizing the deviation of $\frac{m_a}{4\pi}\sum_{nl}(2l+1)|a_{nl}|^2R_{nl}^2(r)/\rho_\mathrm{in}(r)$
from unity. Suppose that the adopted number of $(n,l)$ combinations is $M$. We then select $N \gg M$ radial points $\{r_1,r_2,\cdots,r_N\}$ within $1\,\mathrm{kpc}$. The problem can be cast as a least-squares minimization
\begin{equation}
    \min_{\boldsymbol{A}}||X_{N\times M}\boldsymbol{A}_{M}-\boldsymbol{I}_{N}||^2,
\end{equation}
where $\boldsymbol{I}$ is an $N$-dimensional unit vector. The elements of the matrix $X$ and the vector $\boldsymbol{A}$ are defined as 
\begin{equation}
X_{ij}=\frac{m_a(2l_j+1)}{4\pi}\frac{R_{n_jl_j}^2(r_i)}{\rho_\mathrm{in}(r_i)}, \quad A_j=|a_{n_jl_j}|^2,
\end{equation}
where $i=1,2,\cdots, N$ and $j=1,2,\cdots, M$, and $n_j$ and $l_j$ denote the values of $n$ and $l$ in the $j$-th $(n,l)$ pair. Unlike a standard least-squares problem, the elements of $\boldsymbol{A}$ represent squared expansion coefficients and must therefore be non-negative. We thus solve this problem using a non-negative least-squares algorithm implemented in the SciPy library.

\section{Constraints of the parameter space\label{Sec4}}

For each parameter set $(m_a, f_a)$, we construct the FDM wavefunction at the present time $t_0$ using the median density profile obtained from the Jeans analysis as $\rho_{\mathrm{in}}(r)$. From this wavefunction, we derive the corresponding FDM density profile $\rho_{\mathrm{out}}(r) = m_a|\psi(r)|^2$ and evaluate its consistency with the input profile. This comparison allows us to determine whether the model parameters are consistent with the observational data \cite{Zimmermann:2024xvd}.

\subsection{Statistics}
We introduce a statistical measure to quantify the agreement between $\rho_{\mathrm{out}}(r)$ and $\rho_{\mathrm{in}}(r)$, thereby deriving constraints on $(m_a, f_a)$. In this study, we employ three methods to set these constraints.

\begin{figure}[htbp]
    \centering    
    \includegraphics[width=\linewidth]{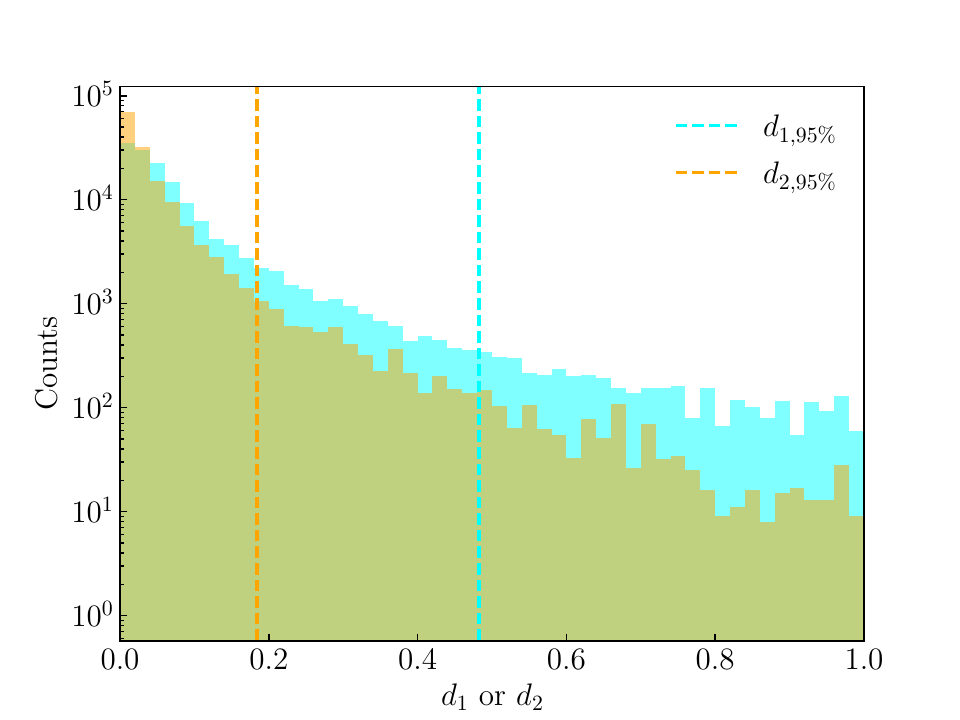}
    \caption{Distribution of the distances $d_1$ and $d_2$ between the $1.596\times 10^5$ density profiles obtained from the Jeans analysis and the median profile $\rho_\mathrm{in}$. The cyan and orange dashed lines indicate the corresponding 95th percentiles, above which $5\%$ of the samples lie.}
    \label{hist}
\end{figure} 

The first two methods rely on a distance measure between the two density profiles. Specifically, for a radial interval $[r_\mathrm{min},r_\mathrm{max}]$, we introduce a symmetric distance between two arbitrary density profiles $\rho_1$ and $\rho_2$ as 
\begin{equation}
\begin{aligned}
    &d(\rho_1,\rho_2;r_\mathrm{min},r_\mathrm{max})\equiv\frac{1}{\ln(r_\mathrm{max}/\mathrm{kpc})-\ln(r_\mathrm{min}/\mathrm{kpc})}\\
    &\times \frac{1}{2}\int_{r_\mathrm{min}}^{r_\mathrm{max}}\left\{\left[\frac{\rho_1(r)}{\rho_2(r)}-1\right]^2+\left[\frac{\rho_2(r)}{\rho_1(r)}-1\right]^2\right\}d\ln\frac{r}{\mathrm{kpc}},
\end{aligned}
\end{equation}
where the logarithmic radial variable is adopted to prevent the large-$r$ region from dominating the integral. The distance measures used in the first two methods are then defined as 
\begin{equation}
\begin{aligned}
&d_1(\rho_1,\rho_2)\equiv d(\rho_1,\rho_2;r_{1/2}/4,4r_{1/2}),\\
&d_2(\rho_1,\rho_2)\equiv d(\rho_1,\rho_2;r_{1/2}/4,r_{1/2}), 
\end{aligned}
\end{equation}
respectively. The difference between these two methods lies in the radial interval used for the constraint. The first method adopts the same radial range as that used in Ref.~\cite{Zimmermann:2024xvd}, namely $[r_{1/2}/4,4r_{1/2}]$, corresponding to the yellow shaded region in Fig.~\ref{fit}. The second method instead restricts the comparison to the inner part of this region, i.e., the range inside the half-light radius $r_{1/2}$ indicated by the black dashed line.

We determine whether a given parameter set can be accepted at the $95\%$ confidence level as follows. Taking $d_1$ as an example, we first compute the distances $\{d_1(\rho_{\mathrm{in}}, \rho_i),i=1,2,\cdots,1.596 \times 10^5\}$, where $\rho_i$ denote the profiles obtained from the Jeans analysis. The resulting distribution of $d_1$ is shown in Fig.~\ref{hist}. From this distribution, we determine a $95$-th percentile $d_{1,95\%}$, shown by the cyan dashed line in Fig.~\ref{hist}, such that $5\%$ of the $d_1$ values exceed $d_{1,95\%}$. If the distance $d_1(\rho_{\mathrm{in}}, \rho_{\mathrm{out}}) $ computed for a given parameter set $(m_a,f_a)$ exceeds $d_{1,95\%}$, that parameter set is considered excluded at the $95\%$ confidence level.

The third method is not based on a distance measure. Instead, at each radius within the interval $[r_{1/2}/4,4r_{1/2}]$, we determine the value corresponding to the lowest $5\%$ of the density distribution obtained from the Jeans analysis. These values are then connected to form the black curve shown in Fig.~\ref{fit}. In the third method, a model is considered consistent with the data at the $95\%$ confidence level only if $\rho_{\mathrm{out}}(r)$ remains entirely above this curve throughout the interval.

\subsection{Results}

\begin{figure}[htbp]
    \centering    
    \includegraphics[width=\linewidth]{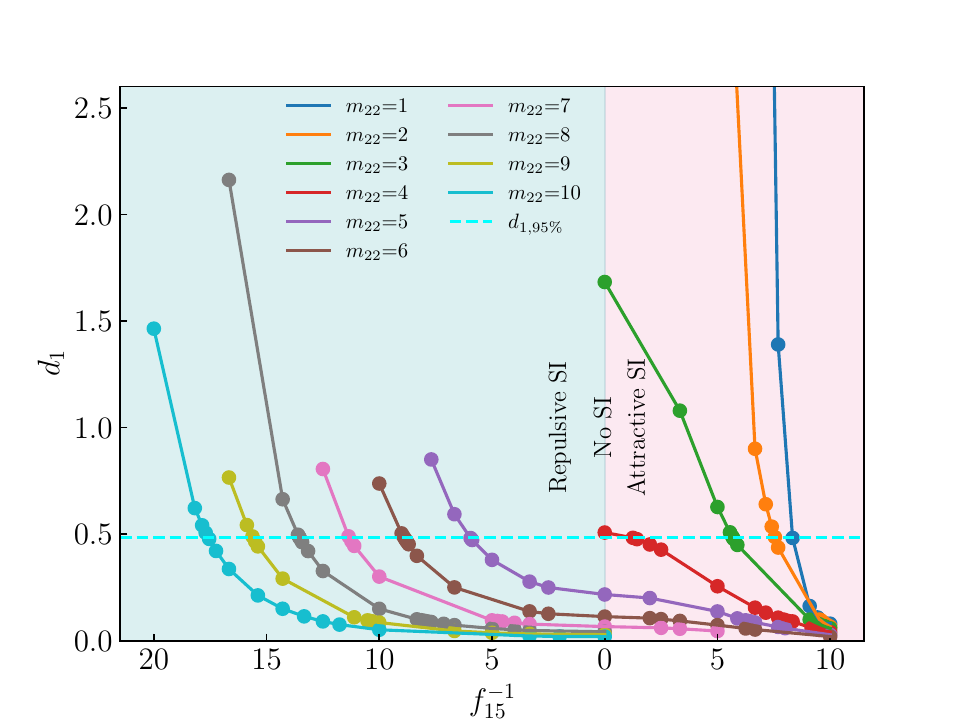}
    \includegraphics[width=\linewidth]{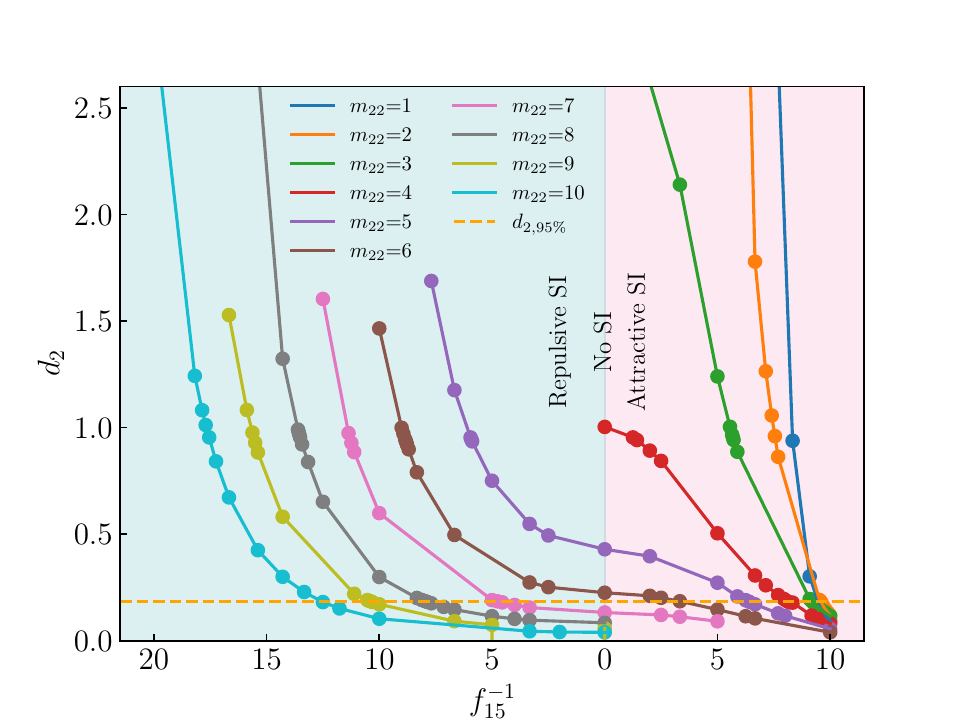}
    \caption{Values of $d_1(\rho_\mathrm{in},\rho_\mathrm{out},)$ (upper panel) and $d_2(\rho_\mathrm{in},\rho_\mathrm{out},)$ (lower panel) as functions of the sign and strength of the SI for different $m_{22}$. The cyan and pink shaded regions correspond to the parameter space of repulsive and attractive SIs, respectively. The boundary between them represents the non-interacting case. The horizontal cyan and orange dashed lines in the upper and lower panels indicate the $95\%$ exclusion thresholds $d_{1,95\%}$ and $d_{2,95\%}$, respectively.}
    \label{d_m_f}
\end{figure} 

For convenience, we introduce two dimensionless quantities:
\begin{equation}
    m_{22}\equiv \frac{m_a}{10^{-22}\mathrm{\, eV}},\quad f_{15}\equiv \frac{f_a}{10^{15}\mathrm{\, GeV}}.
\end{equation}
We first illustrate qualitatively how variations in $m_{22}$ and $f_{15}$ affect the agreement with the results of the Jeans analysis.

In the upper panel of Fig.~\ref{fit}, we show the fitting results for different values of $m_{22}$ in the absence of SI, i.e., $f_{15}=\infty$. As $m_{22}$ decreases, the deviations from the Jeans analysis results become increasingly pronounced. On one hand, a smaller $m_a$ leads to stronger quantum pressure, which enlarges the ground-state solution (the soliton) and produces a large core in the inner halo density profile. Such a core is difficult to reconcile with the relatively cuspy profile inferred from the Jeans analysis for Leo II. On the other hand, enhanced quantum effects associated with a smaller $m_a$ increase the energy spacing between eigenstates, thereby reducing the total number of available eigenstates. This effectively decreases the number of fitting coefficients $a_{nlm}$, reducing the freedom to fit the input profile and consequently worsening the fit, consistent with the discussion in Ref.~\cite{Zimmermann:2024xvd}.

\begin{figure}[htbp]
    \centering    
    \includegraphics[width=0.85\linewidth]{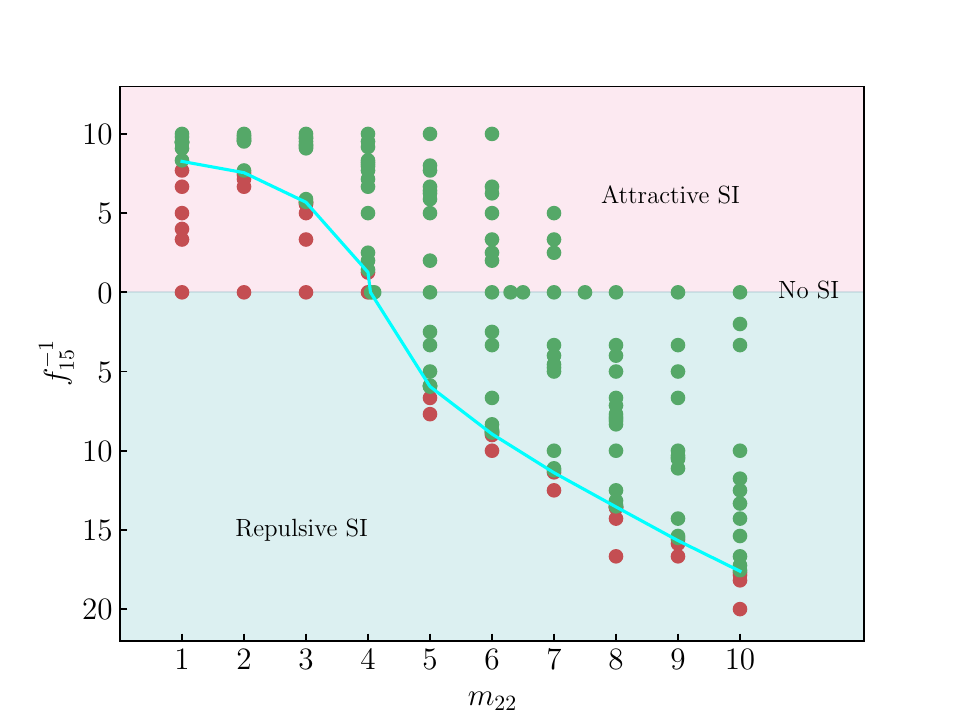}
    \includegraphics[width=0.85\linewidth]{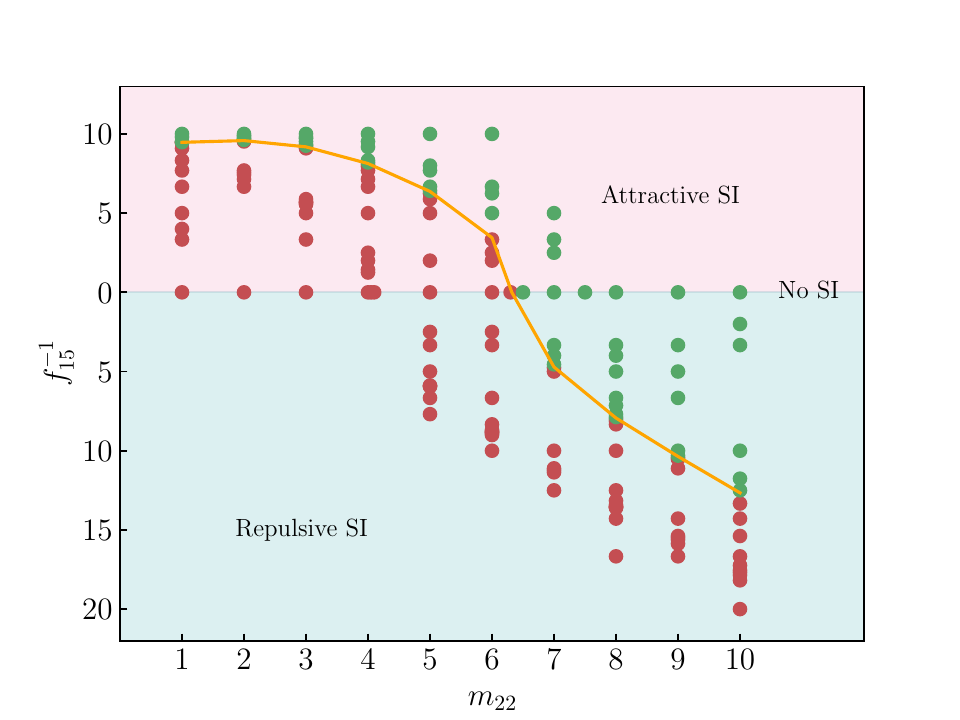}
    \includegraphics[width=0.85\linewidth]{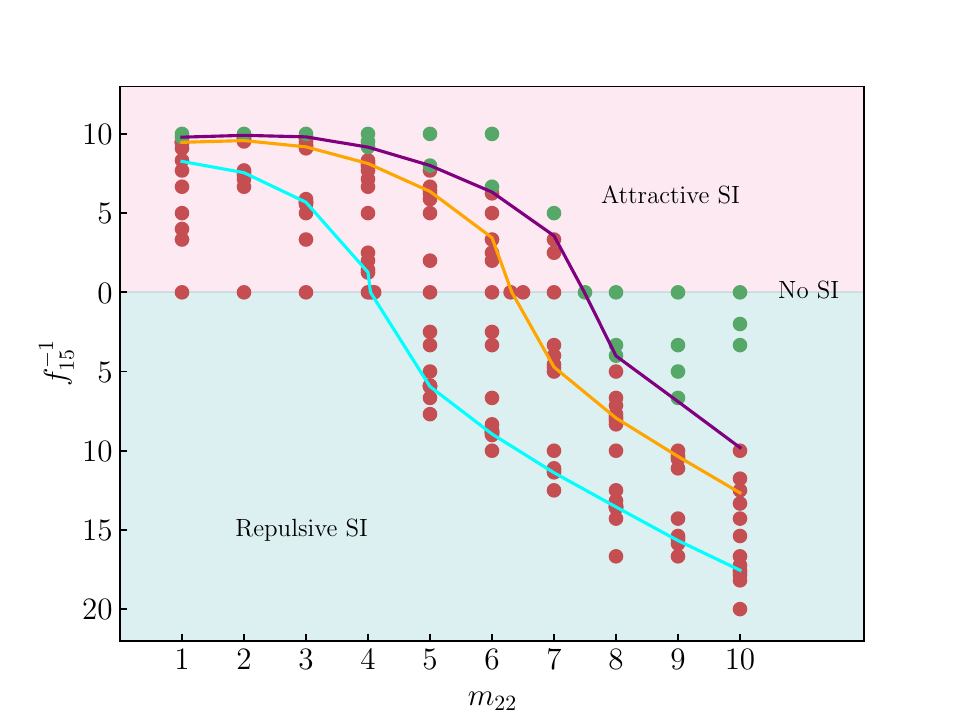}
    \caption{Constraints on the SIFDM parameter space $(m_{22},f_{15})$.
    The upper, middle, and lower panels show the constraints derived from the three statistical methods, represented by cyan, orange, and purple solid curves, respectively. For comparison, the constraints from the first two methods are also displayed in the lower panel. The green and red points denote parameter sets that are accepted and excluded at the $95\%$ confidence level, respectively. The shading of the SI regions follows the same scheme as in Fig.~\ref{d_m_f}.}
    \label{result}
\end{figure} 

In the absence of SI, the $95\%$ confidence-level lower limits on $m_{22}$ obtained from our three statistical methods are $4.1$, $6.3$, and $7.5$, respectively. The difference between the first and the second methods is that the latter places greater emphasis on the quality of the fit at smaller radii. Since the density profile inferred from the Jeans analysis becomes increasingly cuspy toward smaller radii, making it more difficult to reproduce, the second method is expected to yield a stronger constraint than the first.  Our constraints are consistent with those in Ref.~\cite{Zimmermann:2024xvd} at the order-of-magnitude level, although all three of our limits are weaker by a factor of a few. We attribute this difference to variations associated with the statistical method, as also reflected in the differences among the constraints obtained from our three methods.

In the lower panel of Fig.~\ref{fit}, we take $m_{22}=6$ as an example to illustrate the effects of attractive and repulsive SIs of varying strengths on the fitting results. As the attractive (repulsive) SI strengthens, the agreement with the Jeans analysis improves (worsens) relative to the non-interacting case. This behavior arises because attractive (repulsive) SI reduces (increases) the soliton size and correspondingly raise (lower) the central density. Consistently, the upper and lower panels of Fig.~\ref{d_m_f} show the dependence of $d_1(\rho_{\mathrm{in}}, \rho_{\mathrm{out}})$ and $d_2(\rho_{\mathrm{in}}, \rho_{\mathrm{out}})$ on the SI strength $f_{15}^{-1}$ for different values of $m_{22}$, as computed in the first two statistical methods. As the SI varies from strongly repulsive to strongly attractive, both $d_1$ and $d_2$ initially decrease rapidly and then gradually level off. The cyan and orange dashed lines indicate the $95\%$ thresholds $d_{1,95\%}$ and $d_{2,95\%}$, respectively. It can be seen that for larger $m_{22}$, a region of parameter space with repulsive SI remains allowed, whereas for smaller $m_{22}$, sufficiently strong attractive SI is required to avoid exclusion.

We summarize the constraints on the parameter space $(m_{22},f_{15})$ in Fig.~\ref{result}. The three panels display the constraints obtained from the three statistical methods, respectively. Red and green points denote parameter sets that are excluded and accepted at the $95\%$ confidence level, respectively. The cyan, orange, and purple solid curves represent the corresponding $95\%$ exclusion boundaries derived from the three methods. For a more direct comparison of the relative strength of these constraints, the exclusion boundaries from the first two methods are also shown in the third panel. As can be seen, the constraints obtained from the three statistical approaches exhibit noticeable differences.

\section{Conclusion\label{Sec5}}
In this study, we investigate the impact of SI on FDM by confronting theoretical requirements of density profiles with the stellar kinematic constraints from the dwarf spheroidal galaxy Leo II. Our analysis demonstrates that SI plays a qualitatively distinct and physically transparent role in shaping the inner halo structure. For a fixed FDM particle mass, attractive SI leads to a more compact soliton core and an enhanced central density, thereby improving agreement with the relatively cuspy density profiles inferred from Jeans analysis. In contrast, repulsive SI yields more extended cores and suppresses central densities, exacerbating the discrepancy with the kinematic data. This behavior reflects the competition between quantum pressure and SIs.

In the absence of SI, the $95\%$ confidence-level lower limits on $m_{22}$ from the three statistical methods are 4.1, 6.3, and 7.5, respectively. This range underscores the sensitivity of dynamical constraints to the relative weighting of inner versus outer halo regionsprofiles exhibiting stronger cusps at small radii are inherently more challenging to reconcile within the FDM framework. When SI is introduced, the viable parameter space is significantly modified: for relatively large particle masses, both weakly repulsive and attractive interactions remain allowed, whereas for smaller masses, sufficiently strong attractive SI is required to maintain consistency with the data. More generally, we find that for SI strengths satisfying $f_a^{-1}\lesssim 10^{-14}\,\mathrm{GeV}^{-1}$, the allowed range of particle masses lies broadly within $(1-10)\times10^{-22}\,\mathrm{eV}$, although the precise boundaries remain method dependent.

\acknowledgments
This work is supported by the National Natural Science Foundation of China under Grants No. 12575113 and No. 12447105.


\appendix
\section{Jeans analysis based on a  soliton+NFW profile\label{App_A}}

In the framework of FDM, the extremely light
bosonic particles exhibit pronounced wave-like behavior on kiloparsec scales,
leading to distinct dynamical features compared to CDM.
In particular, the inner regions of FDM halos are expected to form a solitonic core, corresponding to the ground-state
solution of the coupled Schr\"{o}dinger--Poisson equations.

High-resolution numerical simulations of FDM halos \cite{2014NatPh..10..496S}
have shown that the radial density profile of the soliton core can be
well approximated by the parametric form
\begin{equation}
\rho_{sol}(r)=\frac{\rho_c}{\left[ 1+0.091(r/r_c)^2 \right]^8},
\label{eq:rho_sol}
\end{equation}
where $r_c$ denotes the core radius and $\rho_c$ is the central density given by
\begin{equation}
\rho_c=1.9\times 10^7 \left(\frac{m_a}{10^{-22}\,{\rm eV}}\right)^{-2}
\left(\frac{r_c}{\rm kpc}\right)^{-4}
\,[M_{\odot} \, {\rm kpc}^{-3}],
\label{eq:rho_c}
\end{equation}
with $m_a$ representing the mass of the FDM particle.

Although the enclosed mass profile can be obtained by direct integration of  Eq.~\eqref{eq:rho_sol}, we employ an accurate approximation for the mass profile to improve computational efficiency in the MCMC analysis \cite{2017MNRAS.468.1338C}:
\begin{equation}
\begin{split}
M&_{\rm{sol}}(r) = \frac{4.2 \times 10^4 \, M_{\odot}}{(m_a/10^{-22} \, \rm{eV})^2 (r_c / \rm{kpc})(a^2 + 1)^7} \\
\bigl(&3465a^{13} + 23100a^{11} + 65373a^9 + 101376a^7 + 92323a^5 \\
&+ 48580a^3 - 3465a + 3465(a^2 + 1)^7 \arctan(a)\bigr),
\end{split}
\label{eq:analy_mass}
\end{equation}
where $a=(2^{1/8}-1)^{1/2}(r/r_c)$.

Cosmological simulations further reveal a scaling relation between
the soliton core and the host halo, often referred to as the
soliton--halo relation. In particular, the core radius $r_c$ can be
approximately related to the particle mass $m_a$ and the halo
virial mass $M_{200}$ as
\begin{equation}
r_c \approx 1.6 \left(\frac{m_a}{10^{-22}\,{\rm eV}}\right)^{-1}
\left(\frac{M_{200}}{10^{9}M_{\odot}}\right)^{-1/3}
\,{\rm kpc},
\label{eq:sol_r_c}
\end{equation}
which reflects the inverse dependence of the core size on both the
particle mass and the halo mass.

Outside the central soliton, the density profile transitions to a
standard NFW-like form, establishing a composite soliton+NFW structure.
The matching between the inner soliton profile and the outer NFW halo
is defined at a transition radius $r_{\epsilon}$, determined by the
condition
\begin{equation}
\frac{\rho_s}{(r_{\epsilon}/r_s)(1+r_{\epsilon}/r_s)^2}
=
\frac{\rho_c}{\left[1+0.091(r_{\epsilon}/r_c)^2\right]^8}
=
\epsilon \rho_c,
\label{eq:epsilon_rho_c}
\end{equation}
where $\epsilon$ parametrizes the matching density in units of the
central soliton density. This construction ensures a smooth connection
between the inner and outer regions of the halo. Numerical simulations \cite{2014NatPh..10..496S,2018PhRvD..97h3519M} suggest that the
transition radius typically satisfies $r_{\epsilon} \gtrsim 3r_c$,
which we adopt as a physically motivated constraint in our analysis.

A concentration--mass relation derived from CDM
simulations is imposed as an additional constraint in the analysis.
Such a relation encapsulates the correlation between
halo mass and concentration.
The concentration is modeled as
\begin{equation}
\begin{split}
C_{200}(M_{200},x_{\rm sub}) = &c_{0}\left[1+\sum_{i=1}^3
\left(a_i{\rm log}_{10}\left(\frac{M_{200}}{10^8 h^{-1} M_{\odot}} \right)\right)^i \right] \\
&\times \left[1+b \, {\rm log}_{10}(x_{\rm sub}) \right],
\end{split}
\label{eq:C_M_x}
\end{equation}
where $c_0=19.9$, $a_i=(-0.195,0.089,0.089)$, and $b=-0.54$ are
the best-fit parameters reported in Ref.~\cite{2017MNRAS.466.4974M}, and $M_{200}$ is derived from Eq.~\ref{eq:sol_r_c}. The dimensionless parameter $x_{sub}$ is defined as the ratio of the
subhalo's distance to the host halo center relative to the host virial
radius:
\begin{equation}
x_{\rm sub}=\frac{r_{\rm sub}}{r_{200,{\rm MW}}},
\end{equation}
with $r_{200,{\rm MW}}=210 \,{\rm kpc}$ adopted throughout this analysis.

On the other hand, the concentration parameter can be directly
determined by the density profile obtained from the Jeans analysis. To enforce consistency between this
concentration $c_{200}$ and the simulation expectation, we consider an additional
likelihood term \cite{Hayashi:2021xxu}:
\begin{equation}
-2\log\mathcal{L}_{c}=\frac{\left[{\rm log}_{10}(c_{200})-{\rm log}_{10}(C_{200})\right]^2}{\sigma_{\rm CDM}^2},
\label{eq:likelihood_c}
\end{equation}
where $\sigma_{\rm CDM}=0.13$ \cite{2017MNRAS.466.4974M} denotes the intrinsic scatter of the relation. The total likelihood is then constructed as the sum of the kinematic
likelihood and the concentration prior,
\begin{equation}
\mathcal{L}_{\rm tot}=\mathcal{L}_{\rm dsph}+\mathcal{L}_{c}.
\end{equation}

In this framework, the free parameters of the model are taken to be
$m_a$, $r_c$, $\epsilon$, $r_s$, and $\beta$.
We adopt uniform priors over the following ranges:
$-3<{\rm log}_{10}(m_a/10^{-23}\,{\rm eV})<8$,
$-7<{\rm log}_{10}(r_c/{\rm kpc})<5$,
$-5<{\rm log}_{10}(\epsilon)<{\rm log}_{10}(0.5)$,
$-3<{\rm log}_{10}(r_s/{\rm kpc})<1$, and
$-5<\beta<1$.
We perform the MCMC sampling using $1.035\times10^6$ samples.
The resulting posterior distributions yield
${\rm log_{10}}(m_a/10^{-23}{\rm eV})=3.2_{-0.9}^{+2.6}$,
${\rm log_{10}}(r_c/{\rm kpc})=-2.5_{-1.9}^{+1.2}$,
${\rm log_{10}}(\epsilon)=-3.4_{-1.0}^{+0.9}$,
${\rm log_{10}}(r_s/{\rm kpc})=0.1_{-0.4}^{+0.2}$,
and $\beta=-0.4_{-2.8}^{+0.7}$.

\bibliography{Refs}

\end{document}